\documentclass[onecolumn]{emulateapj}
\usepackage{natbib}
\shortauthors{Chung \& Ryu}
\shorttitle{ microlensing events by wide separation planets with a moon}
\newcommand{\te}{t_{\rm E}}

\newcommand{\thetae}{\theta_{\rm E}}

\newcommand{\thetaep}{\theta_{\rm E,p}}
\newcommand{\qm}{q_{\rm m}}
\newcommand{\sm}{s_{\rm m}}
\newcommand{\mpp}{m_{\rm p}}
\newcommand{\qp}{q_{\rm p}}
\newcommand{\spp}{s_{\rm p}}
\newcommand{\delxm}{\Delta x_{\rm m}}
\newcommand{\delxmj}{\Delta x_{\rm m, J}}
\newcommand{\delcs}{\Delta \chi^{2}}
\begin{document}
\title{Properties of microlensing events by wide separation planets with a moon}
\author{
Sun-Ju Chung\altaffilmark{1}
and 
Yoon-Hyun Ryu\altaffilmark{1}
 }
\altaffiltext{1}
{Korea Astronomy and Space Science Institute 776, Daedeokdae-ro,
Yuseong-Gu, Daejeon 34055, Republic of Korea; sjchung@kasi.re.kr}
\altaffiltext{2}
{http://exoplanet.eu/catalog}

% ==================================================================
%\submitted{Submitted to The Astrophysical Journal}

\begin{abstract}
We investigate the properties of microlensing events caused by planetary systems where planets with a moon are widely separated from their host stars.
From this investigation, we find that the moon feature generally appears as an very short-duration perturbation on the smooth asymmetric light curve of the lensing event induced by the wide separation planet; thus it can be easily discriminated from the planet feature responsible for the overall asymmetric light curve.
For typical Galactic lensing events with the Einstein radius of $\sim 2$ AU, the asymmetry of the light curves due to bound planets can be noticed up to $\sim 20$ AU.
We also find that the perturbations of the wide planetary systems become dominated by the moon as the projected star-planet separation increases, and eventually the light curves of events produced by the systems appear as the single lensing light curve of the planet itself with a very short-duration perturbation induced by the moon, which is a representative light curve of the event induced by a star and a planet, except on the Einstein timescale of the planet.
We also study the effect of a finite source star on the moon feature in the wide planetary lensing events.
From this study, we find that when the lunar caustic is sufficiently separated from the planetary caustic, the lower limit on the ratio of the lunar caustic size to the source radius causing a $\geq 5\%$ lunar deviation depends mostly on the projected planet-moon separation regardless of the moon/star mass ratio, and it decreases as the planet-moon separation becomes smaller or larger than the planetary Einstein radius.

\end{abstract}

\section{INTRODUCTION}

Microlensing is a very powerful method for detecting exoplanets with wide separations from host stars, whereas most of about 2000 exoplanets discovered so far are hot-Jupiter planets with close separations, which were detected by radial velocity and transit methods.
Microlensing is sensitive to not only wide separation planets but also to low-mass planets (Bennett \& Rhie 1996; Beaulieu et al. 2006).
Hence, microlensing is one of the most powerful methods for studying planet formation and evolution.

The microlensing signal of a planet is a short-duration perturbation on the smooth standard light curve of the lensing event induced by the host star that occurred on a background source star.
The planetary perturbation is induced by two sets of disconnected caustics that are composed of the central and planetary caustics.
When the central and planetary caustics are merged at $s \sim 1$, the caustic is called the resonant caustic, where $s$ is the projected separation of the star and the planet normalized by the Einstein radius of the lens system, i.e.,
\begin{equation}
\thetae = \sqrt{{ {4GM \over {c^2}} \left({1\over D_L} - {1\over D_S}\right)}},
\end{equation}
where $M$ is the total mass of the lens system, and $D_{\rm L}$ and $D_{\rm S}$ are the distances to the lens and the source from the observer, respectively.
The central caustic always produces central perturbations around the peak of the lensing light curve because of the formation of the central caustic close to the host star, whereas the planetary caustic produces perturbations at any part of the light curve because it is formed away from the star.
For the detection of the planetary perturbations, survey and follow-up microlensing observations are being carried out toward the Galactic bulge.
The survey observations (OGLE: Udalski 2003; MOA: Bond et al. 2002) monitor a large field of sky and alert ongoing events with and without anomalies by analyzing data in real time, while the follow-up observations ($\mu$FUN: Dong et al. 2006; PLANET: Albrow et al. 2001: RoboNet: Burdorf et al. 2007; MiNDSTEp: Dominik et al. 2010) densely monitor the alerted events.
As a result, 43 exoplanets have been discovered so far by microlensing \altaffilmark{2}.

Recently, studies on wide separation planetary lensing events have been reported (Chung \& Lee 2011; Sumi et al. 2011; Ryu et al. 2013; Freeman et al. 2015).
In particular, \citet{sumi11} reported that a population of unbound or distant massive planets is about two times larger than that of main-sequence stars.
This result implies the importance of detection of events induced by wide separation planets and free-floating planets.
However, it is not easy to detect those events due to the short event duration (e.g., a few days for a Jupiter-mass planet).
To detect the events, it is necessary to have 24 hr continuous and high cadence observations together with high photometric accuracy for a wide field of view.
In 2015, Korea Microlensing Telescope Network (KMTNet) has initiated a 24 hr continuous observation toward the Galactic bulge using a 1.6 m wide-field telescope at each of three different sites, Chile, South Africa, and Australia \citep{kim10}.
KMTNet has been observing four fields with a total field of view of $4\arcdeg \times 4\arcdeg$, in which each field has a field of view of $2\arcdeg \times 2\arcdeg$ and has a cadence of about 10 minutes \citep{kim10}.
In addition, space-based microlensing observations with high cadence and high photometric accuracy, such as the \textit{Wide-Field InfraRed Survey} (\textit{WFIRST}) \citep{spergel15} and the \textit{Euclid} \citep{penny13} missions, will be carried out toward the Galactic bulge in the near future.
Thanks to the advanced ground- and space-based observations, it is expected to detect plentiful wide separation planets and free-floating planets with masses of down to Earth-mass or below. 

Considering the Solar system, most planets would have one or more moons.
In the Solar system, although the number of moons of planets differ, the ratio of the total mass of the moons to the planet mass is $\lesssim 10^{-4}$ for all the planets in the system, except for the Earth with the Moon and the Pluto with the Charon, which have ratios of $0.012$ and $\sim 0.1$, respectively \citep{canup06}.
From this, we can infer that the ratio of the moon mass to the planet mass would be commonly very low, and thus for planets under the strong lensing effect of a host star (e.g., planets located within a lensing zone of $0.6 \lesssim s \lesssim 1.6$), in general, it is very difficult to detect the moon signal, because the lunar caustic is extremely small compared to  the planetary caustic and thus perturbations induced by the lunar caustic are buried by strong finite-source effects and planetary caustic perturbations.
However, as the star-planet separation increases, the lensing effect of the host star rapidly decreases and the planetary caustic also shrinks rapidly.
Hence, planetary systems with wide separations of $s \geq 3$ are much more sensitive to the moon signal, even though the probability of occurring events of the wide separation planets is lower than that for events of planets located in the lensing zone.
In this case, the planetary caustic can be similar to or smaller than the lunar caustic, and thus it is important to distinguish between the planet and moon signals. 
Moreover, since the ground- and space-based observations are expected to yield more detections of wide separation planets and free-floating planets, it is important to study the lensing properties of events by wide separation planetary systems that have a moon and understand the lunar and planetary perturbations.
\citet{han02} studied the feasibility of detecting moons of exoplanets with microlensing and found that it will be very difficult to detect the moon signals because of the severe finite-source effect.
\citet{han08} studied planetary systems where a planet hosting a massive moon is located within the lensing zone of $0.6 \lesssim s \lesssim 1.6$ and found that Earth-mass moons may be detectable when the distance from the planet is similar to or greater than the Einstein radius of the planet.
\citet{liebig10} also studied the detectability of moons around exoplanets with microlensing and showed that massive moons can be detected in principle via the technique of Galactic microlensing.
Since all those studies considered only planets located within the lensing zone, studies on the lensing effects of moons in planetary systems with wider separations are needed.
In this paper, we investigate the properties of events caused by planetary systems where planets with wide separations of $s \geq 3$ host a moon.

This paper is organized as follows.
In Section 2, we briefly describe the multiple lensing.
In Section 3, we investigate perturbations of events caused by wide separation planetary systems with a moon and compare them with perturbations of events caused by the planetary systems without a moon.
We also investigate the change of the lunar caustic perturbations on finite-source effects and the photometric accuracy for the detection of the lunar perturbations in events of dwarf stars in the Galactic bulge.
In Section 4, we discuss possible planetary systems to mimic events caused by wide separation planetary systems with and without a moon.
We summarize the results and conclude in Section 5.

\section{MULTIPLE LENSING}

For multiple lens systems composed of $N$-point masses, the lens equation \citep{witt90} is expressed as
\begin{equation}
\zeta = z - \sum_{j=1}^{N}{m_{j}/M\over{\overline{z} - \overline{z}_{j}}},
\end{equation}
where $\zeta = \xi + i\eta$, $z_{j} = x_{j} + iy_{j}$, and $z = x + iy$ represent the complex notations of the source, lens, and image positions, respectively, $z_j$ and $m_j$ are the position and mass of the $j$th lens, $M$ is the total mass of the lens system, and $\overline{z}$ and $\overline{z}_j$ denote the complex conjugates of $z$ and $z_j$.
The locations of images for given positions of the lens and source are obtained by the inversion of the lens equation.
Magnifications $A_i$ of the individual images $z_i$ are given by the inverse of the determinant of the Jacobian of the lens equation evaluated at each image position (Gaudi \& Gould 1997; Han et al. 2001; Liebig \& Wambsganss 2010)
\begin{equation}
A_{i} = \left(1\over{|{\rm det}J|}\right)_{z=z_i} ;\\\ {\rm det}J = 1 - {\partial\zeta\over{\partial\overline{z}}}{\overline{\partial\zeta}\over{\partial\overline{z}}}.
\end{equation}
The total magnification is the sum of those of the individual images, i.e., $A = \sum_{i}A_i$.
$|{\rm det}J| = 0$ means that the magnification of a point source is infinite, and the set of positions in the source plane on which the magnification of the point source is infinite is a caustic.
The caustic is formed only in multiple lensing ($N \geq 2$), and thus it is the most different feature of the multiple lensing compared to the single lensing.

For a triple lens system composed of a host star, a planet, and a moon, the lensing behavior can be described by the superposition of those of two planetary lens systems composed of the star-planet and star-moon pairs, when the planet and moon are located far outside the Einstein radii of the star and planet, respectively (Han et al. 2001; Han \& Han 2002; Han 2008).
In planetary lensing, the effective lensing position of the planet is shifted toward the host star and the amount of the shift of the planet (Di Stefano \& Mao 1996; An \& Han 2002; Chung  et al. 2005) is 
\begin{equation}
\Delta x_{\rm shift, p \rightarrow s} = \left({{m_{\rm s}/m_{\rm p}}\over{s}}\right)\left(\thetaep\over{\thetae}\right)^{2},
\end{equation}
where $m_{\rm p}$ and $m_{\rm s}$ are the masses of the planet and star and $\thetaep$ is the Einstein radius of the planet mass.
The planetary caustic is formed at the shifted position of $s - \Delta x_{\rm shift, p \rightarrow s}$.
The shape and number of the planetary caustic depend primarily on the star-planet separation, while the size of the caustic depends on the planet/star mass ratio as well.
For wide separation planets ($s > 1$), the planetary caustic appears as a single \textbf{astroid-shaped }caustic, where the horizontal size on the star-planet axis is generally longer than the vertical size \citep{chung11}.
As the separation increases, the caustic becomes symmetric.
The horizontal and vertical sizes of the planetary caustic \citep{han06} are expressed as
\begin{equation}
\Delta x_{\rm w} \simeq {4\sqrt{q}\over s^{2}}\left (1 + {1 \over 2s^{2}}\right )\\\ {\rm and} \\\
\Delta y_{\rm w} \simeq {4\sqrt{q} \over s^{2}}\left (1 - {1 \over 2s^{2}}\right ) .
\end{equation}

For close separation planets ($s < 1$), the planetary caustic appears as two same triangular-shaped caustics, and the two caustics are symmetrically displaced perpendicular to the star-planet axis.
In this case, $s - \Delta x_{\rm shift}$ produces a $`` (-)"$ value, and it means that the caustics are located at the opposite side of the star from the planet.
The horizontal and vertical sizes of the twin caustic \citep{han06} are expressed as
\begin{equation}
\Delta x_{\rm c} \simeq {{3\sqrt{3} \over 4} \sqrt{q}s^{3}} \\\ {\rm and} \\\
\Delta y_{\rm c}\simeq \sqrt{q}s^{3}.\
\end{equation}
Since $\Delta x_{\rm c} \propto s^{3}$ for close planets, while for wide planets, $\Delta x_{\rm w} \propto s^{-2}$, the caustic of close planets shrinks more rapidly than that of wide planets as the separation becomes farther from 1.
The distance between the two caustics \citep{han06} is represented by
\begin{equation}
h = {4\sqrt{q}(1-s^{2})^{1/2} \over s}.
\end{equation}

\section{PERTURBATIONS BY WIDE SEPARATION PLANETS WITH A MOON}

\subsection{\it Features of planet and moon}

We consider triple lens systems where planets with wide separations of $s \geq 3$ from stars host a moon.
To show perturbations induced by the star and the moon on the position of the planet, we construct the map of the fractional deviation, $\delta$, which is defined as
\begin{equation}
\delta = {A - A_{0} \over {A_0}}\ ,
\end{equation}
where $A$ and $A_0$ are respectively the triple lensing magnification and the single lensing magnification of the planet alone.

Figure 1 shows the fractional deviation maps for triple lens systems composed of a host star, a wide separation planet, and a moon, as a function of the projected star-planet separation, $\spp$, and the projected planet-moon separation, $\sm$, where $\spp$ and $\sm$ are normalized by the Einstein radii of the total lens mass ($\thetae$) and the planet mass ($\thetaep$), respectively.
The maps in the figure are computed by the inverse ray-shooting method where a large number of light rays are uniformly shot backward from the observer plane through the lens plane and then collected in the source plane (Schneider \& Weiss 1986; Wambsganss et al. 1990).
Here we consider the case where the moon is located within the lensing zone of the planet, i.e., $0.6 \lesssim \sm \lesssim 1.6$, in which the lunar caustic is located within the Einstein ring of the planet and thus the probability of detecting the lunar perturbations is high.
In the figure, brown and blue color regions represent the positive and negative deviation areas, and the color changes into darker scales when the deviation is $|\delta| = 1\%,\ 5\%,\ 10\%,\ 15\%,\ 30\%$, and $60\%$, respectively.
Considering of typical Galactic lens systems, we assume that the mass and distance of the lens are $M_{\rm L} = 0.3\ M_{\odot}$, $D_{\rm L} = 6\ \rm{kpc}$, respectively, and the distance of the source is $D_{\rm S} = 8\ \rm{kpc}$, and the source is a main-sequence star with a radius of $R_\star = 1.0\ R_\odot$.
Then, the angular and physical Einstein radii of the lens are $\thetae = 0.32\ \rm{mas}$ and $1.91$ AU, and the source radius normalized to the Einstein radius is $\rho_{\star} = \theta_\star/\thetae = (R_\star/D_{\rm S})/\thetae = 0.0018$.
For limb darkening, we adopt a brightness profile for the source star of the form, i.e.,
\begin{equation}
{I(\theta)\over{I_0}} = {1 - \Gamma \left(1-{3\over{2}}{\rm cos}\theta \right) }
\end{equation}
where $\Gamma$ is a limb darkening coefficient and $\theta$ is the angle between the normal to the surface of the source star and the line of sight \citep{an02a}.
We assume that $\Gamma$ is 0.5 for all stars in our simulations.
In Figure 1, the masses of the planet and moon are $\mpp = 1M_{\rm Jupiter}$ and  $m_{\rm m} = 2M_{\rm Mars}$, respectively, which correspond to the planet/star mass ratio of $q_{\rm p} = 3.0\times10^{-3}$ and the moon/star mass ratio of $q_{\rm m} = 2.0\times10^{-6}$. 
The position angle of the moon measured from the star-planet axis is $\phi = 60^{\circ}$.
The black curve in the corner of each map represents the Einstein radius of the planet mass.
The black curve within the Einstein radius represents the triple lensing caustic.
The straight lines with an arrow represent the source trajectories.
Figure 2 shows the resulting light curves and residuals for the source trajectories drawn in Figure 1.
In the figure, the black and red solid curves represent the light curves of triple and binary lensing events with and without a moon, respectively, and the gray dashed curve represents the single lensing light curve of the planet itself.
As one may expect, perturbations induced by a star (or planetary perturbations) are dominant in the Einstein ring of the planet, as seen in Figure 1, even though the planetary caustic is smaller than the lunar caustic (see the second row of panels from bottom).
From this figure, we find that the lensing effect of the star on the planet reaches out to $\spp \simeq 10$, and thus for typical Galactic lensing events with the Einstein radius of $\sim 2$ AU, the boundness of the planet can be noticed up to $\sim 20$AU.
This has been also pointed out by \citet{han09}.
The dominant planetary perturbations give rise to the overall asymmetric light curves of the triple lensing events, as shown in Figure 2.
Then the asymmetric light curve indicates that the planet is bound to the star.
The star-planet separation that can notice the boundness of the planet increases with an increase of the planet mass and a decrease of the source diameter \citep{ryu13}.

From Figures 1 and 2, we find that the moon feature generally appears as a very short-duration perturbation on the smooth asymmetric light curves induced by the wide separation planet, and thus it can be easily discriminated from the planet feature responsible for the asymmetric light curve.
We also find that perturbations of the wide planetary systems with a moon become dominated by the moon as the star-planet separation increases, and eventually the light curves of events produced by the systems appear as the standard single lensing light curve of the planet itself with a very short-duration perturbation induced by the moon.
The resulting light curve is a representative light curve of the event induced by a star and a planet, except on the Einstein timescale of the planet. 
This is because even though the planetary caustic rapidly decreases with an increase of the star-planet separation, the lunar caustic depends primarily on the planet-moon separation and the moon/planet mass ratio.
As a result, if single lensing events of $\te < 2$ days with a very short-duration perturbation are detected, the very short-duration perturbation would be induced by the moon.
However, we cannot insist that the very short-duration perturbation is a moon feature, because the short-duration single lensing light curve with a very short-duration perturbation can be produced by the star-planet lensing event caused by a fast moving lens star or a high velocity source star, as shown in the result of \citet{bennett14}.
Figure 3 shows the best-fit binary lensing light curves for triple lensing events marked as I, II, and III in Figure 2.
As shown in the figure, for two cases of $\spp = 10.0$ and $\spp = 12.0$, the triple and best-fit binary lensing light curves are indistinguishable, unlike the case of $\spp = 7.0$.
Hence, this figure shows that the triple lensing events become well matched with the binary lensing events as the star-planet separation increases.

To show only perturbations induced by a moon, we construct the fractional deviation maps between wide separation planetary lens systems with and without the moon, i.e., triple and binary lens systems, in Figure 4.
From this figure, we find that for wide planetary lens systems with $\spp \gtrsim 3$ the effect of the star-planet separation on the lunar perturbations is weak.
This means that the star-planet separation for the wide planetary lens systems is responsible for the overall asymmetric light curve, but it is not mostly concerned with the lunar perturbations.

\subsection{\it Finite-source effect }

For wide planetary lens systems with $\spp \geq 3$, the finite-source effect is severe because the size of the planetary caustic is comparable to the diameter of the source star.
Hence, the limb darkening of the finite source surface should be considered for the lens systems.
Here we adopt the same limb darkening coefficient used in Figures $1 - 4$.

Figure 5 shows the fractional deviation maps between wide planetary lens systems with and without a moon for various ratios of the lunar caustic size to the source radius, $\delxm/\rho_\star$.
The lensing parameters of the triple lens systems are $\spp = 4.0$, $\qp = 3.0\times10^{-3}$, $\qm = 2.0\times10^{-6}$, and $\phi = 60^\circ$.
The lunar caustic size $\delxm$ is defined as the distance between two cusps on the planet-moon axis.
The red circle in the map represents the source star and its size $R_\star$ is presented in each map.
From the figure, we find that the ratio $\delxm/\rho_\star$ causing a $\geq 5\%$ lunar deviation decreases as $\sm$ becomes farther from 1.
This means that the perturbation induced by the lunar caustic becomes stronger as the caustic shrinks with the increase (for $\sm > 1$) or the decrease (for $\sm < 1$) of $\sm$.
This is because for planetary lens systems composed of a star and a planet, the positive perturbation of the planetary caustic becomes stronger as the caustic shrinks, while the negative perturbation becomes weaker \citep{chung11}.
Thanks to this property of the planetary caustic, the positive perturbation of the lunar caustic becomes also stronger as the caustic shrinks, as shown in Figure 5 (see also Figure 4).
Figure 5 also shows that for cases that the lunar caustic is combined with the planetary caustic at $\sm \simeq 1.0$, the lunar caustic perturbation becomes weaker by interacting with the planetary caustic perturbation, and thus the $\geq 5\%$ lunar deviation occurs when the ratio $\delxm/\rho_\star$ is relatively much bigger than the cases of $\sm > 1$ and $\sm < 1$.
We also estimate the lower limit on the ratio $\delxm/\rho_\star$ causing the $\geq 5\%$ lunar deviation for various $\qm$, as a function of $\sm$, and the result is presented in Figure 6.
Here we assume that $\qp = 3 \times 10^{-3}$, $\spp = 4.0$, and $\phi = 60^\circ$.
Two dotted vertical lines indicate the positions at $\sm =0.9$ and $\sm =  1.15$, and the region between the two lines represents the region where the lunar resonant caustic is formed.
From Figure 6, we find that when the lunar caustic is sufficiently separated from the planetary caustic, the lower limit of the ratio $\delxm/\rho_\star$ causing the $\geq 5\%$ lunar deviation depends mostly on $\sm$ regardless of $\qm$.
This implies two facts.
First, if $\sm$ is the same, the lunar caustic perturbation changes with $\qm$ by a factor of $\sqrt{q_{\rm m, 2}/q_{\rm m, 1}}$, which is the increment or the decrement of the caustic size as $q_{\rm m, 1}$ changes to $q_{\rm m, 2}$.
Second, when applying the source radius being multiplied by the factor of $\sqrt{q_{\rm m, 2}/q_{\rm m, 1}}$, the lunar caustic perturbation pattern of the triple lens system with $q_{\rm m, 2}$ is almost the same as that of the triple lens system with $q_{\rm m, 1}$, except the perturbation area, which means the perturbation duration.
According to Figure 6, the $\geq 5\%$ lunar deviation for $\sm = 1.2$ occurs when $\delxm/\rho_\star \gtrsim 1.2$.
For cases of $\sm \sim 1$, the lower limit of the ratio generally decreases with the increases of $\qm$, even though it also changes with $\sm$.
This is because as $\qm$ increases, the size of the lunar resonant caustic increases and thus the perturbation of the caustic becomes stronger.
Figure 7 shows the fractional deviation maps between wide planetary lens systems with and without a moon for various moon position angles, $\phi$, and planet/star mass ratios, $\qp$.
From the figure, we find that the strength of the lunar caustic perturbation does not generally change with $\phi$ and $\qp$, except cases with $\sm < 1$ and $\phi \lesssim 20^\circ$, while the perturbation area changes with the two parameters.
As shown in the panel of $\sm = 0.8$, the lunar negative perturbation of the triple lens system with $\phi = 5^\circ$ is weaker than those of other triple lens systems.

We estimate the lunar caustic size as a function of the planet-moon separation, $\sm$, and the moon/planet mass ratio, $\qm/\qp$, when it is sufficiently separated from the planetary caustic, and the result is presented in Figure 8.
Here we assume that the planet/star mass ratio and star-planet separation are $\qp = 3.0\times10^{-3}$ and $\spp = 4.0$, and the moon position angle measured from the star-planet axis is $\phi = 60^{\circ}$
As mentioned before, since the lunar caustic perturbation depends primarily on $\sm$ and $\qm/\qp$, we do not consider the change of the lunar caustic size with $\spp$ and $\phi$.
Although the lunar resonant caustic is combined with the planetary caustic, it is visible well, and thus it can be separately estimated from the planetary caustic.
In Figure 8, the horizontal dotted line indicates the mass ratio of $\qm/\qp = 3.3\times10^{-4}$, which corresponds to the mass ratio between $1M_{\rm Mars}$ moon and $1M_{\rm Jupiter}$ planet.
From Figures 6 and 8, we find that for $\gtrsim 1M_{\rm Mars}$ moons,  a $\geq 5\%$ lunar deviation in events of dwarf stars occurs in most of the lensing zone of the planet, i.e., $0.06{\rm AU} \lesssim d_{\rm m} \lesssim 0.17 \rm AU$, where $d_{\rm m}$ is the physical planet-moon separation.
This means that the $\gtrsim 1M_{\rm Mars}$ moons could potentially be detectable if the dwarf source star crosses or passes close to the lunar caustic.
For the moon/planet mass ratio of $< 10^{-4}$, such as the Jupiter and the Ganymede, there is no region that can occur the $\geq 5\%$ lunar deviation, as shown in Figure 8.
This is because for $\sm \simeq 1.0$ the $\geq 5\%$ lunar deviation occurs when $\delxm/\rho_{\star} \gtrsim 4.0$, as shown in the case of $\qm = 5.0\times10^{-7}$ of Figure 6.
Thus, for typical Galactic bulge events with $\rho = 0.0018$, it seems impossible to detect moons for planetary systems with $\qm/\qp < 10^{-4}$.
If the moon/planet mass ratio and the planet-moon separation normalized by the Einstein radius of the planet are the same, the lunar caustic size normalized by the planetary Einstein radius is also the same.
From this fact and Figure 8, one can estimate the lunar caustic size for other planet masses, and it is represented by
\begin{equation}
{\delxm = \delxmj \sqrt{{\qp\over{3.0\times 10^{-3}}}}},
\end{equation}
where $\delxmj$ is the lunar caustic size for a Jupiter-mass planet with $\qp = 3.0\times10^{-3}$ used in Figure 8.
Equation (10) can be used only in the limit of distant planets ($\spp \gg  1$)  and sufficiently separated lunar caustics.

We also investigate the photometric accuracy for the detection of the lunar deviation in events of dwarf stars with $1.0\ R_\odot$ in the Galactic bulge.
For this, we examine how many photons are needed to detect the lunar deviation in the events, and use the typical lens and source parameters used in Section 3.1.
Here we test only some events caused by a single triple lens system where the moon position angle is $\phi = 60^{\circ}$.
We assume that the extinction toward the Galactic bulge is $A_{I} = 1.0$ and the blended flux is equivalent to the flux of the background star with the apparent magnitude of $I = 20.0$.
We assume that the apparent magnitude of the source star affected by the assumed extinction is $I = 19.0$ and the Einstein timescale is $\te = 20$ days, and the observation cadence is once per 15 minutes, having the exposure time of 1 minute.
We assume that the total surface brightness from the dark sky and the Moon is $I = 19.7$ mag/arcsec$^2$ \citep{henderson14}, and the readout noise and image scale are 10 $e^-$ and 0.4 arcsec/pixel, which are the same as those of KMTNet.
Considering that \citet{henderson14} used the planet detection threshold of $\delcs_{\rm th} = 160$, we also assume that the moon is detected only if $\delcs_{\rm t} - \delcs_{\rm b} \geq 160$, where $\delcs_{\rm t}$ and $\delcs_{\rm b} $ are $\delcs$ of triple and binary lensing events from the best-fit single lensing event, respectively.
Since the light curves of the binary lensing events without the moon correspond to those of the triple lensing events with the moon except around the lunar perturbation, especially for the triple lensing events with remarkable asymmetric light curves, as shown in Figure 2, we do not conduct the binary lensing fitting.
In this study, we add Gaussian scatter to the points of the simulated triple lens light curves.
From this investigation, we find that for the detection of the lunar deviations of 5 \%  and 10 \% in events of a $I = 19.0$ dwarf star, it is required to detect at least 84 and 32 photons s$^{-1}$ for an $I = 20$ star, which give the accuracies of 0.7\% and 1.2\% at $I = 18.0$, respectively.
Figure 9 shows example light curves of events of an $I = 19.0$ star with the lunar deviations of $\simeq 7 \%$  and $\simeq 12 \%$, in which the events with the individual photon acquisition rates of 84 and 32 photons s$^{-1}$ satisfy the above lunar detection condition.
We also do the same test for dwarf stars with different brightness of $I = 18.0$ and $I = 18.5$, and the result is presented in Table 1.
In this test, we assume that the radii of the stars are all $1.0\ R_\odot$.
The photon acquisition rates of KMTNet and OGLE are about 30 photons s$^{-1}$ and about 20 photons s$^{-1}$ for an $I = 20$ star (Chung et al. 2014 ; Henderson et al. 2014), which give the accuracies of $\sim 1.2\%$ and $\sim 1.6\%$ at $I = 18.0$, respectively, and other microlensing monitoring and follow-up telescopes also have typically the accuracy of $\sim 2\%$ \citep{liebig10}.
Considering this fact, the result in Table 1 implies that for events of brighter dwarf stars than $I = 19.0$, the $\gtrsim 10 \%$ lunar deviation can be detected from current observational systems, and especially, for events of $I \sim 18.0$ stars it is possible to detect the 5 \% lunar deviation by using the current systems.

\section{DISCUSSION}

As shown in Figure 1, for wide separation planetary systems with a massive planet and a low-mass moon with a moon/planet mass ratio of $6.7\times 10^{-4}$, planetary perturbations are dominant in the Einstein ring of the planet.
Thus, the moon feature appears as an additional very short-duration perturbation on the asymmetric light curve induced by the wide separation planet.
However, in the case of a wide separation low-mass planet with a massive moon, such as a Neptune-mass planet with an Earth-mass moon with a moon/planet mass ratio of $0.1$, the lunar caustic is much bigger than the planetary caustic, and thus the lunar perturbations are dominant in the Einstein ring of the planet (see Figure 10).
This causes events produced by such a lens system to mimic the asymmetry lensing light curve of the planet.
In this case, although the moon-induced perturbations strongly appear in the light curve, it is difficult to discriminate them from the planetary perturbations.

Multiple planetary systems composed of two wide separation planets have a potential to produce perturbations induced by a moon in wide separation planetary systems.
In order to mimic the lunar perturbations, one of the two planets should be located near the Einstein radius of the other planet, but in principle, it seems unlikely that such a planet exists because if they are located close to each other, then they will eventually collide by their mutual gravity.
Even in our solar system, there are no such planets, especially among planets with wide separations of $\geq 5$ AU.
In microlensing, however, the separation between the star and the planet is the projected separation, and thus it is possible that one planet is located near the Einstein radius of the other planet.
However, the probability of such a case occurring is extremely low compared to the planet-moon case.
Assuming that the Solar system and source star are located at 6 kpc and 8 kpc, respectively, and the orbit of planets in the Solar system is edge-on, for the Saturn and the Neptune, it takes individually about 8 and 35 years to enter them into the Einstein ring of the Jupiter, whereas 66 out of 67 moons of the Jupiter locate within the lensing zone of the Jupiter.
Therefore, the probability that the multiple wide separation planetary systems mimic the lunar perturbations is very low.

In typical Galactic events, the Einstein timescales of a Jupiter-mass planet and an Earth-mass planet are roughly $\te = 2 \sim 3$ days and $\te = 2 \sim 3$ hours, respectively.
For a Mars-mass moon, it is roughly $\te = 30 \sim 40$ minutes.
Until now there have been a few of well covered events with $\te < 2$ days  (Sumi et al. 2011; Bennett et al. 2014).
Current microlensing experiments are capable of observing the wide field of the Galactic bulge with high cadence of about $10-20$ minutes by using global network telescopes \citep{skowron15}.
This allows us to detect the short-duration events of $\te < 2$ days caused by a wide separation planet or a free-floating planet, and it also makes it possible to detect a moon feature lasting for $< 1$ hour.
In addition, as mentioned before, considering that KMTNet is currently conducting a 24 hr ground-based observation with high photometric accuracy and high cadence of about 10 minutes and the space-based observations, such as the \textit{WFIRST} and \textit{Euclid} missions, will be sensitive to events due to isolated Mars-mass planetary objects (Spergel et al. 2015; Penny et al. 2013), we carefully expect that the detection of a moon in wide separation planetary systems will be achieved in near future.

\section{CONCLUSION}

We have studied the properties of events caused by a moon in wide separation planetary systems of $\spp \geq 3$.
From the study, we found that the moon feature generally appears as a very short-duration perturbation on the smooth asymmetric light curves induced by the wide separation planet; thus it can be easily discriminated from the planet feature responsible for the overall asymmetric light curve.
We also found that perturbations of the wide planetary systems become dominated by the moon as the star-planet separation increases, and eventually the light curves of the events produced by the systems appear as the single lensing light curve of the planet itself with a very short-duration perturbation induced by the moon, which is a representative light curve of the star-planet event, except on the Einstein timescale of the planet. 
This is because even though the planetary caustic rapidly decreases with an increase of the star-planet separation, the lunar caustic depends primarily on the planet-moon separation and the moon/planet mass ratio.
We have studied the effect of a finite source star on the moon feature in the wide planetary lensing events.
From this study, we found that when the lunar caustic is sufficiently separated from the planetary caustic, the lower limit on the ratio of the lunar caustic size to the source radius causing a $\geq 5\%$ lunar deviation depends primarily on the planet-moon separation regardless of the moon/star mass ratio, and it decreases as the planet-moon separation becomes smaller or larger the planetary Einstein radius.
Finally, we have done a simple investigation about the photometric accuracy for the detection of the lunar deviation in events of dwarf stars with various brightness in the Galactic bulge, which considers only some events caused by a single triple lens system.
As a result, we found that for the detection of the lunar deviations of 5\% and 10\% in events of a $I = 19.0$ dwarf star, it is required the accuracies of at least 0.7\% and 1.2\% at $I = 18.0$, respectively, and for events of brighter dwarf stars than $I = 19.0$, the $\gtrsim 10\%$ lunar deviation can be detected from current observational systems with the accuracy of $\lesssim 2\%$ at $I = 18.0$.

%Figure 1 --------------------------------------------------------------
\begin{figure}[t]
\epsscale{1.0}
\plotone{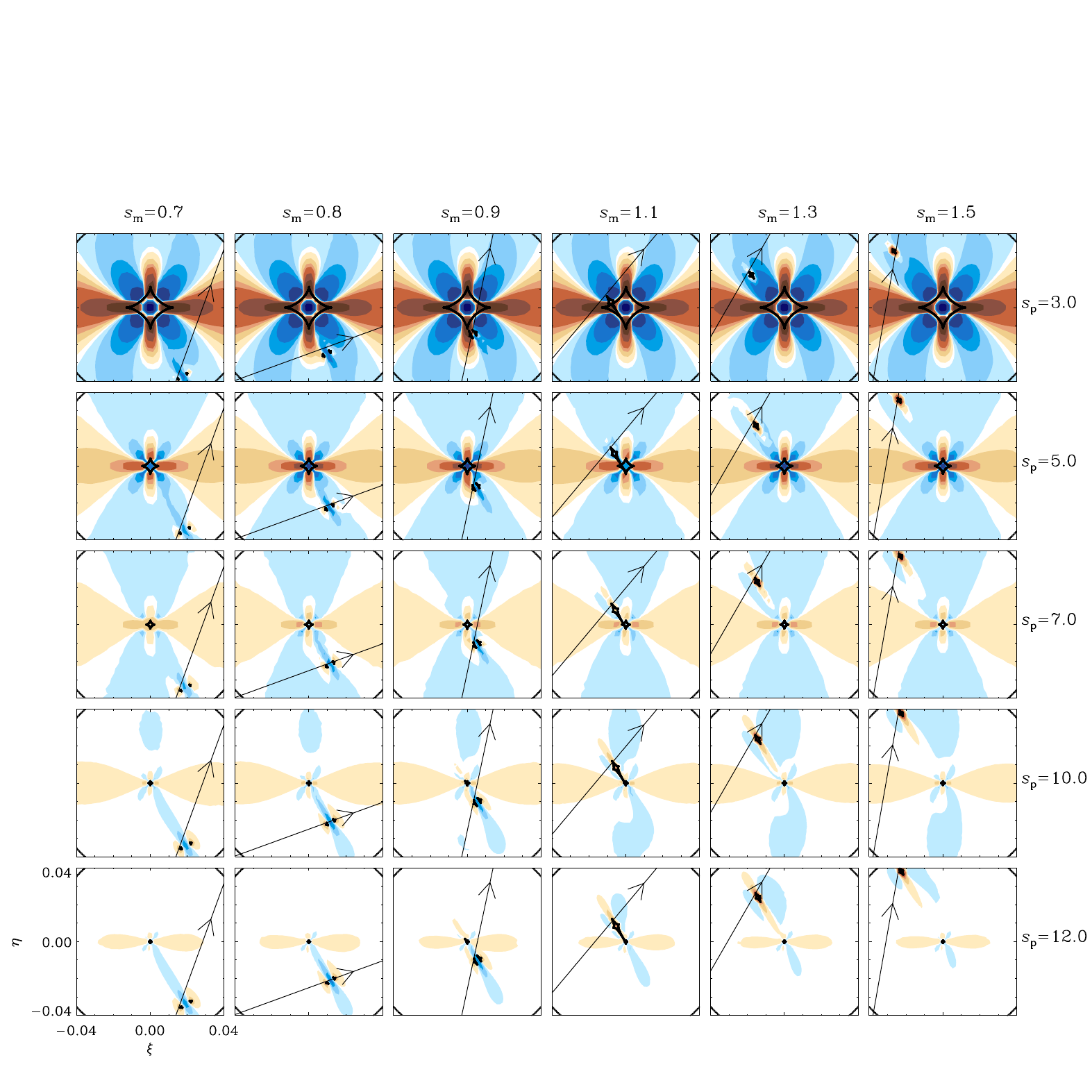}
\caption{\label{fig:one}
Fractional deviation maps between wide separation planetary lens systems with a moon and single lens systems of a planet alone (triple and single lens systems), as a function of the projected star-planet separation, $\spp$, and the projected planet-moon separation, $\sm$, where $\spp$ and $\sm$ are normalized by the Einstein radii of the total lens mass ($\thetae$) and the planet mass ($\thetaep$), respectively.
The coordinates ($\xi$,$\eta$) of the maps are centered at the planetary caustic center.
In the map, the host star is located to the left of the star-planet axis.
Considering of typical Galactic lens systems, we assume that the mass and distance of the lens are $M_{\rm L} = 0.3\ M_{\odot}$, $D_{\rm L} = 6\ \rm{kpc}$, respectively, and the distance of the source is $D_{\rm S} = 8\ \rm{kpc}$, and the source is a main-sequence star with a radius of $R_\star = 1.0\ R_\odot$.
Then, the angular and physical Einstein radii of the lens are $\thetae = 0.32\ \rm{mas}$ and $1.91$ AU, and the source radius normalized to the Einstein radius is $\rho_{\star} = 0.0018$.
In each map, brown and blue color regions represent the positive and negative deviation areas, and the color changes into darker scales when the deviation is $|\delta| = 1\%,\ 5\%,\ 10\%,\ 15\%,\ 30\%$, and $60\%$, respectively.
In the maps, the masses of the planet and moon are $\mpp = 1M_{\rm Jupiter}$ and  $m_{\rm m} = 2M_{\rm Mars}$, which correspond to the planet/star mass ratio of $q_{\rm p} = 3.0\times10^{-3}$ and the moon/star mass ratio of $q_{\rm m} = 2.0\times10^{-6}$, respectively.
The position angle of the moon measured from the star-planet axis is $\phi = 60^{\circ}$.
The black curve in the corner of each map represents the Einstein ring of the planet.
The black curve within the Einstein ring represents the caustic induced by the triple lens systems.
The straight lines with an arrow represent the source trajectories, for which light curves are shown in Figure 2.
}
\end{figure}

%Figure 2 --------------------------------------------------------------
\begin{figure}[t]
\epsscale{1.0}
\plotone{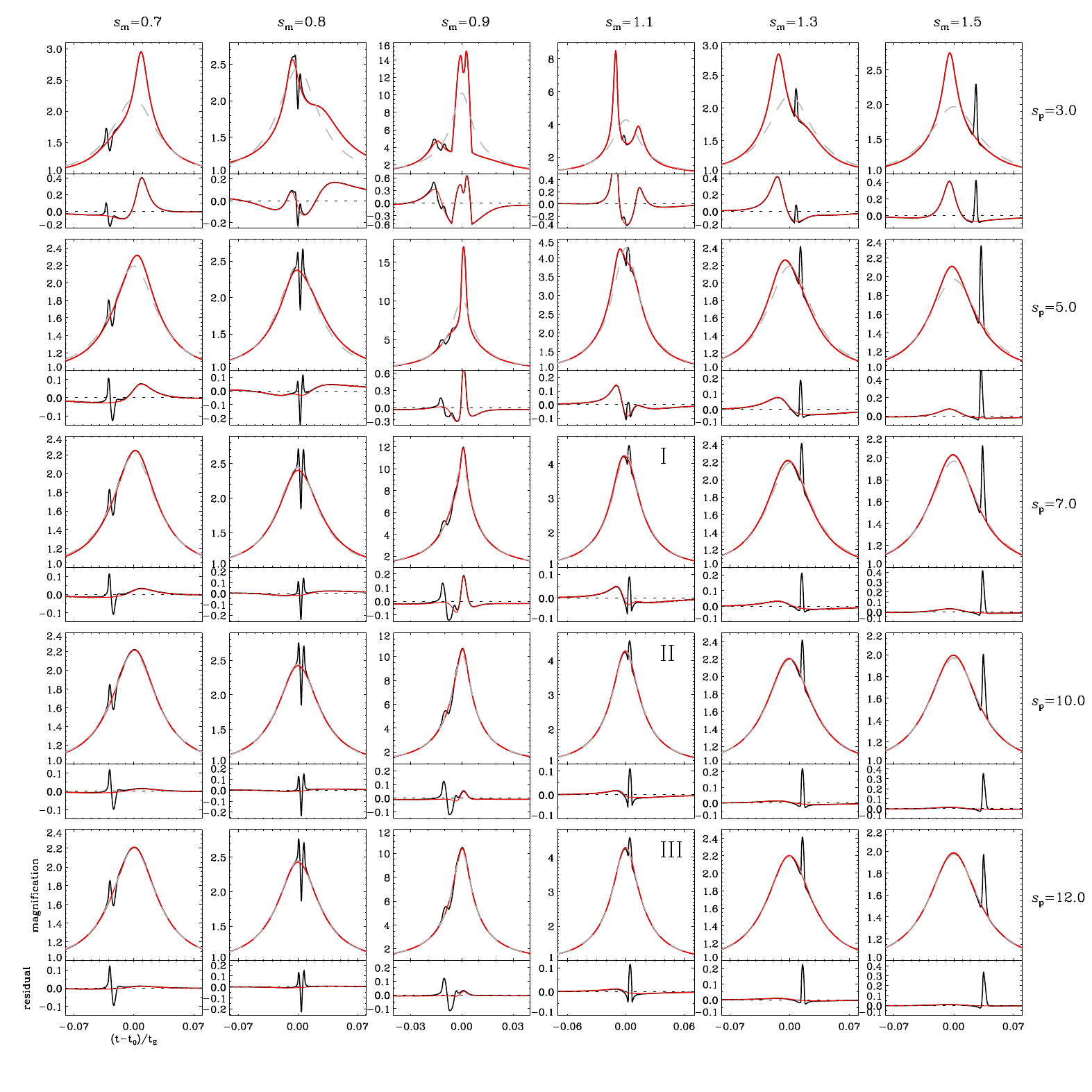}
\caption{\label{fig:two}
Light curves and residuals for the source trajectories presented in Figure 1.
The black and red solid curves represent the triple and binary lensing light curves of the planetary systems with and without a moon, respectively, and the gray dashed curve represents the single lensing light curve of a planet itself.
Residuals are the residuals from the single lensing event.
Triple lensing events with labels I, II, and III are fitted to binary lensing events, and the best-fit result is presented in Figure 3.
}
\end{figure}

%Figure 3 --------------------------------------------------------------
\begin{figure}[t]
\epsscale{1.0}
\plotone{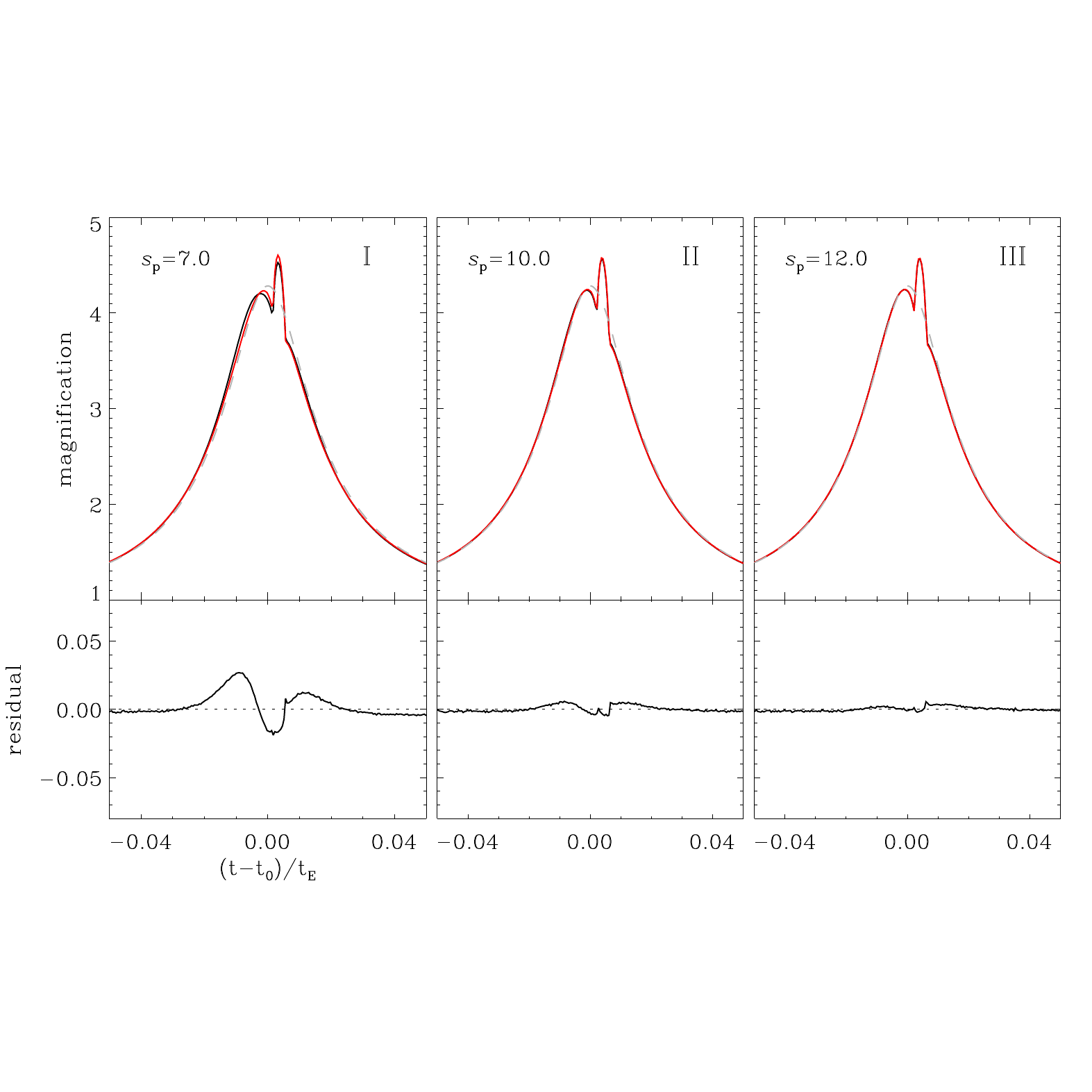}
\caption{\label{fig:three}
Light curves of the best-fit binary lensing events for the triple lensing events marked in Figure 2.
The black and red solid curves are the light curves of the triple lensing and best-fit binary lensing events and the gray dashed curve is the light curve of the planet itself.
The residual represents the difference between the triple and the best-fit binary lensing light curves.
}
\end{figure}

%Figure 4 --------------------------------------------------------------
\begin{figure}[t]
\epsscale{1.0}
\plotone{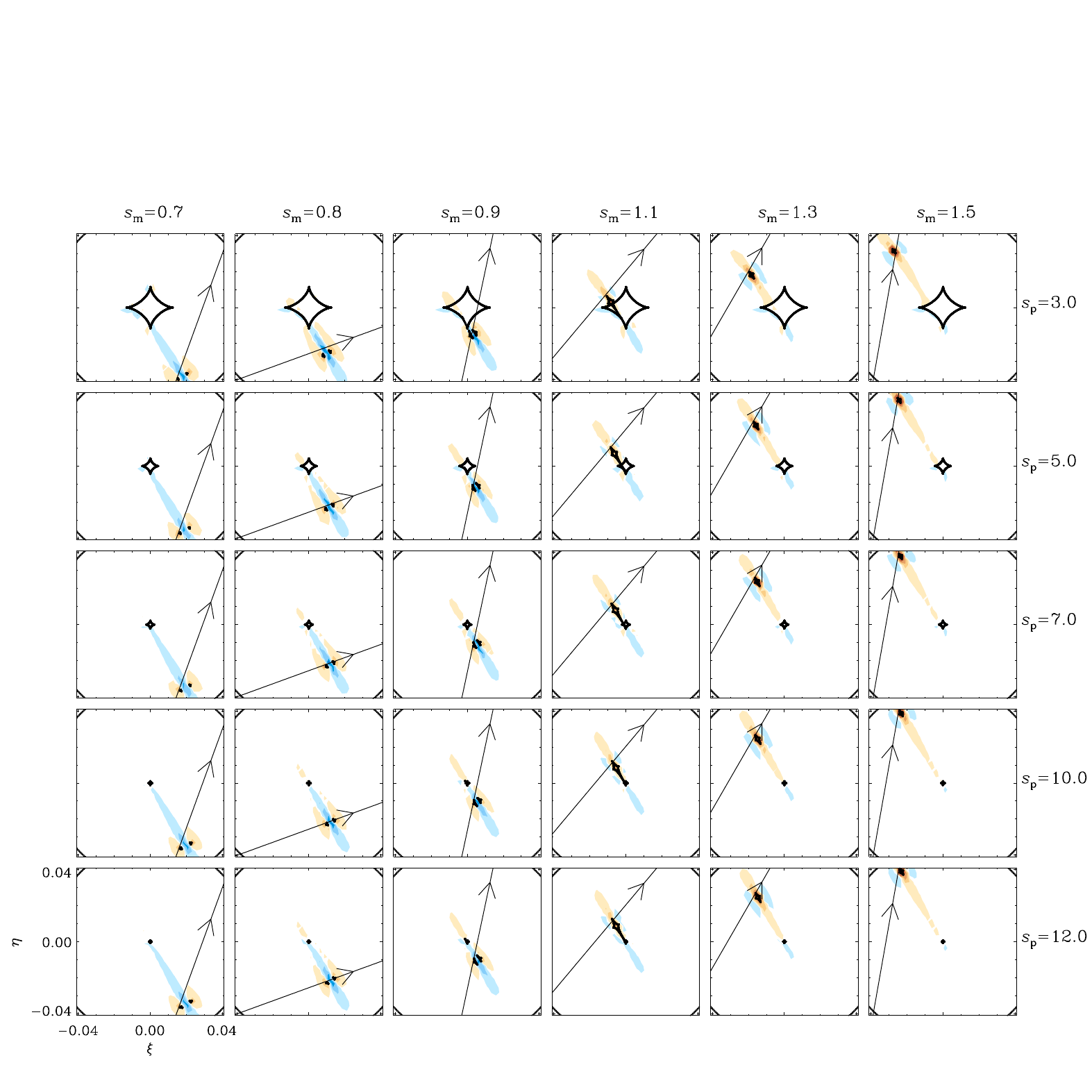}
\caption{\label{fig:four}
Fractional deviation maps between wide separation planetary lens systems with and without a moon (triple and binary lens systems).
This figure shows only the perturbations induced by the moon, whereas Figure 1 shows the perturbations induced by both the star and the moon.
All lensing parameters are the same as Figure 1.
}
\end{figure}

%Figure 5 --------------------------------------------------------------
\begin{figure}[t]
\epsscale{1.0}
\plotone{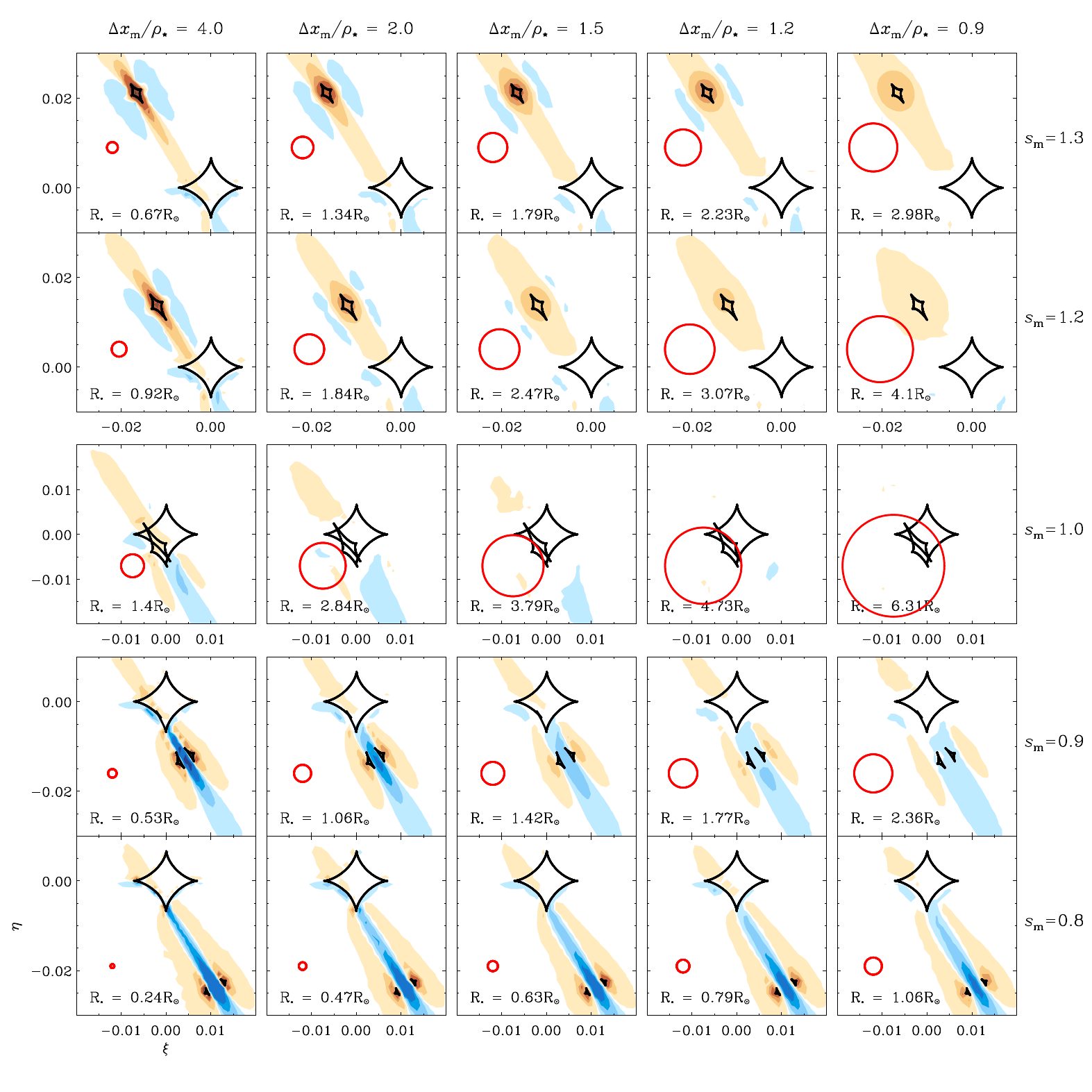}
\caption{\label{fig:five}
Fractional deviation maps between wide separation planetary lens systems with and without a moon for various ratios of the lunar caustic size to the source radius, $\delxm/\rho_\star$.
The lensing parameters of the triple lens systems are $\spp = 4.0$, $\qp = 3.0\times10^{-3}$, $\qm = 2.0\times10^{-6}$, and $\phi = 60^\circ$.
The red circle in the map represents the source star and its size $R_\star$ is presented in each map.
The lunar caustic size $\Delta x_{\rm m}$ is defined as the distance between two cusps on the planet-moon axis.
Brown and blue color regions are the same as those expressed in Figure 1.
}
\end{figure}

%Figure 6 --------------------------------------------------------------
\begin{figure}[t]
\epsscale{1.0}
\plotone{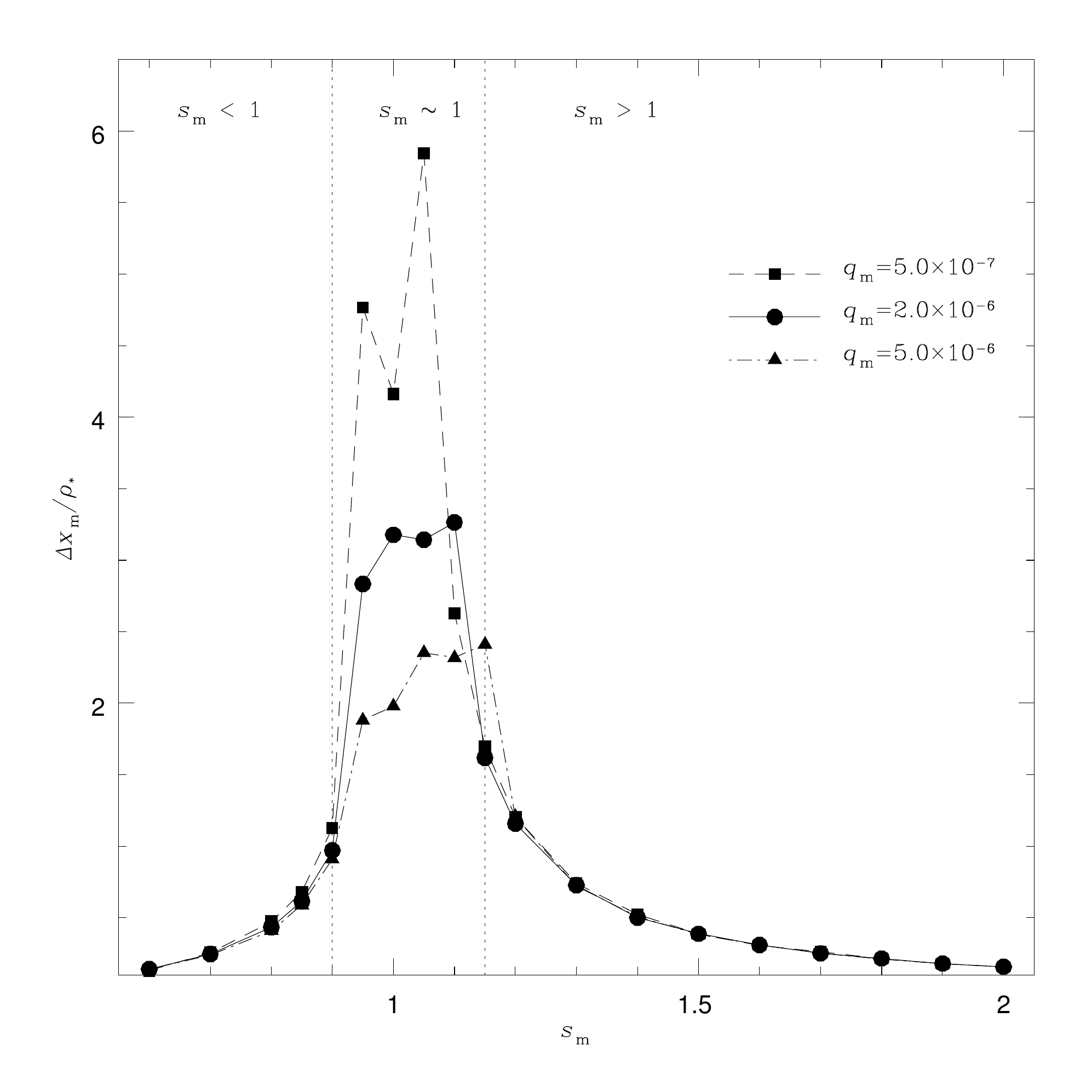}
\caption{\label{fig:six}
Lower limit on the ratio of the lunar caustic size to the source radius, $\delxm/\rho_\star$, causing a $\geq 5\%$ lunar deviation for various moon/star mass ratios, $\qm$, as a function of the planet-moon separation, $\sm$.
Here we assume that $\qp = 3 \times 10^{-3}$, $\spp = 4.0$, and $\phi = 60^\circ$.
Two dotted vertical lines indicate the positions at $\sm =0.9$ and $\sm =  1.15$, and the region between the two lines represents the region where the lunar resonant caustic is formed.
}
\end{figure}

%Figure 7 --------------------------------------------------------------
\begin{figure}[t]
\epsscale{1.0}
\plotone{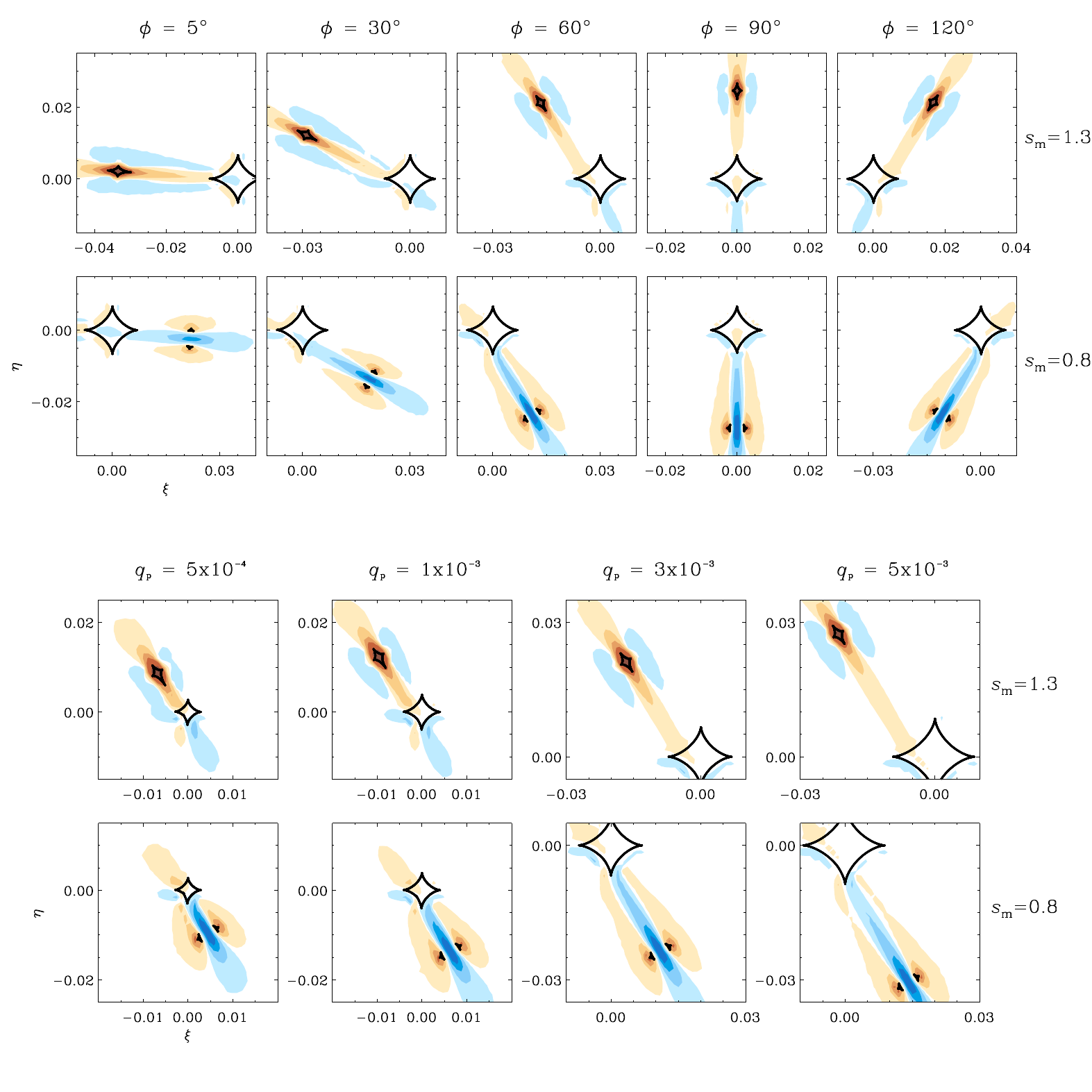}
\caption{\label{fig:seven}
Fractional deviation maps between wide separation planetary lens systems with and without a moon for various moon position angles, $\phi$, and planet/star mass ratios, $\qp$. 
The maps for $\phi$ and $\qp$ are presented in the upper and lower panels, respectively.
The lensing parameters of the triple lens systems in the upper panel are $\qp = 3 \times 10^{-3}$, $\spp = 4.0$, and $\qm = 2.0\times10^{-6}$, while the parameters in the lower panel are $\spp = 4.0$, and $\qm = 2.0\times10^{-6}$, and $\phi = 60^{\circ}$.
The source radius normalized by the assumed typical Einstein radius is $\rho_{\star} = 0.0018$ for both panels.
Brown and blue color regions are the same as those expressed in Figure 1.
}
\end{figure}

%Figure 8 --------------------------------------------------------------
\begin{figure}[t]
\epsscale{1.0}
\plotone{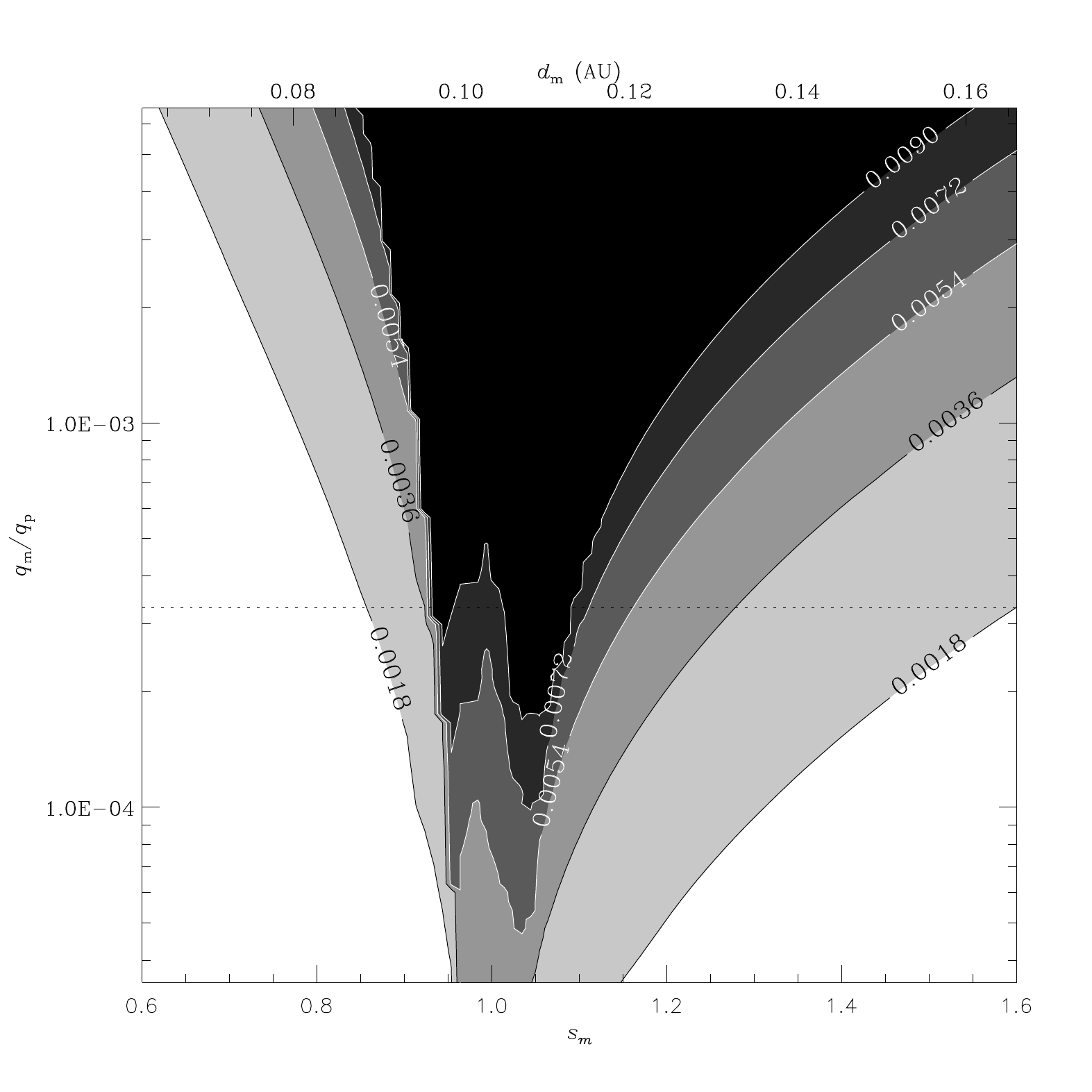}
\caption{\label{fig:eight}
Lunar caustic size $\Delta x_{\rm m}$ as a function of the planet-moon separation, $\sm$, and the moon/planet mass ratio, $\qm/\qp$.
In this calculation, we assume that $\qp = 3 \times 10^{-3}$, $\spp = 4.0$, and $\phi = 60^{\circ}$.
The physical planet-moon separation, $d_{\rm m}$, is also presented.
In the figure, different shades of gray represent the areas with the lunar caustic sizes of $\geq 0.0018$, $\geq 0.0036$, $\geq 0.0054$, $\geq 0.0072 $, and $\geq 0.009$, respectively. 
The contour values were chosen for an easy comparison with source radius 0.0018 used in many of computations.
The horizontal dotted line indicates the mass ratio of $\qm/\qp = 3.3 \times 10^{-4}$, which corresponds to the mass ratio between $1M_{\rm Mars}$ moon and $1M_{\rm Jupiter}$ planet.
}
\end{figure}

%Figure 9 --------------------------------------------------------------
\begin{figure}[t]
\epsscale{1.0}
\plotone{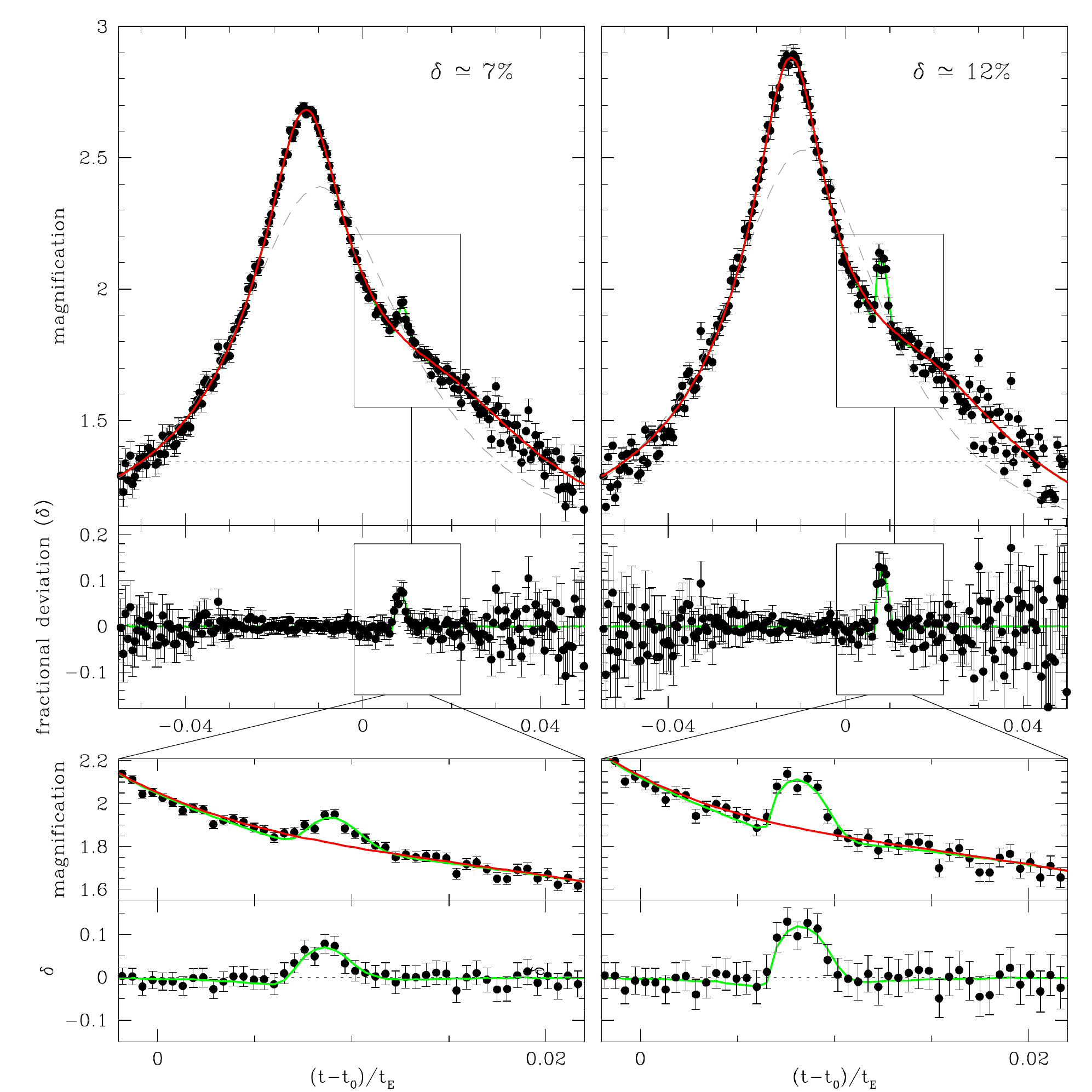}
\caption{\label{fig:nine}
Example light curves of events of an $I = 19$ star with the lunar deviations of $\delta \simeq 7 \%$  and $\delta \simeq 12 \%$, which are included Gaussian scatter to the points of the simulated triple lens light curves.
The photon acquisition rates of 84 and 32 photons $s^{-1}$ for an $I = 20$ star are applied to the events with $\delta \simeq 7 \%$  and $\delta \simeq 12 \%$, respectively. 
As a result, the events with $\delta \simeq 7 \%$ and $\delta \simeq 12 \%$ have $\delcs_{\rm t} - \delcs_{\rm b} = 172.3$ and $\delcs_{\rm t} - \delcs_{\rm b} = 173.6$, respectively, and thus they satisfy the lunar detection condition of $\delcs_{\rm t} - \delcs_{\rm b} \geq 160$, where $\delcs_{\rm t}$ and $\delcs_{\rm b} $ are $\delcs$ of the triple and binary lensing events from the best-fit single lensing event.
The events are magnified by the triple lens system where the masses of the moon and planet are $1M_{\rm Mars}$ moon and $1M_{\rm Jupiter}$, and the projected star-planet and planet-moon separations are  $\spp = 3.0$ and $\sm = 1.3$, respectively, but they have different source trajectories.
The green and red solid curves represent the light curves of the triple and binary lensing events, while the gray dashed curve represents the light curve of the best-fit single lensing event.
}
\end{figure}

%Figure 10 --------------------------------------------------------------
\begin{figure}[t]
\epsscale{1.0}
\plotone{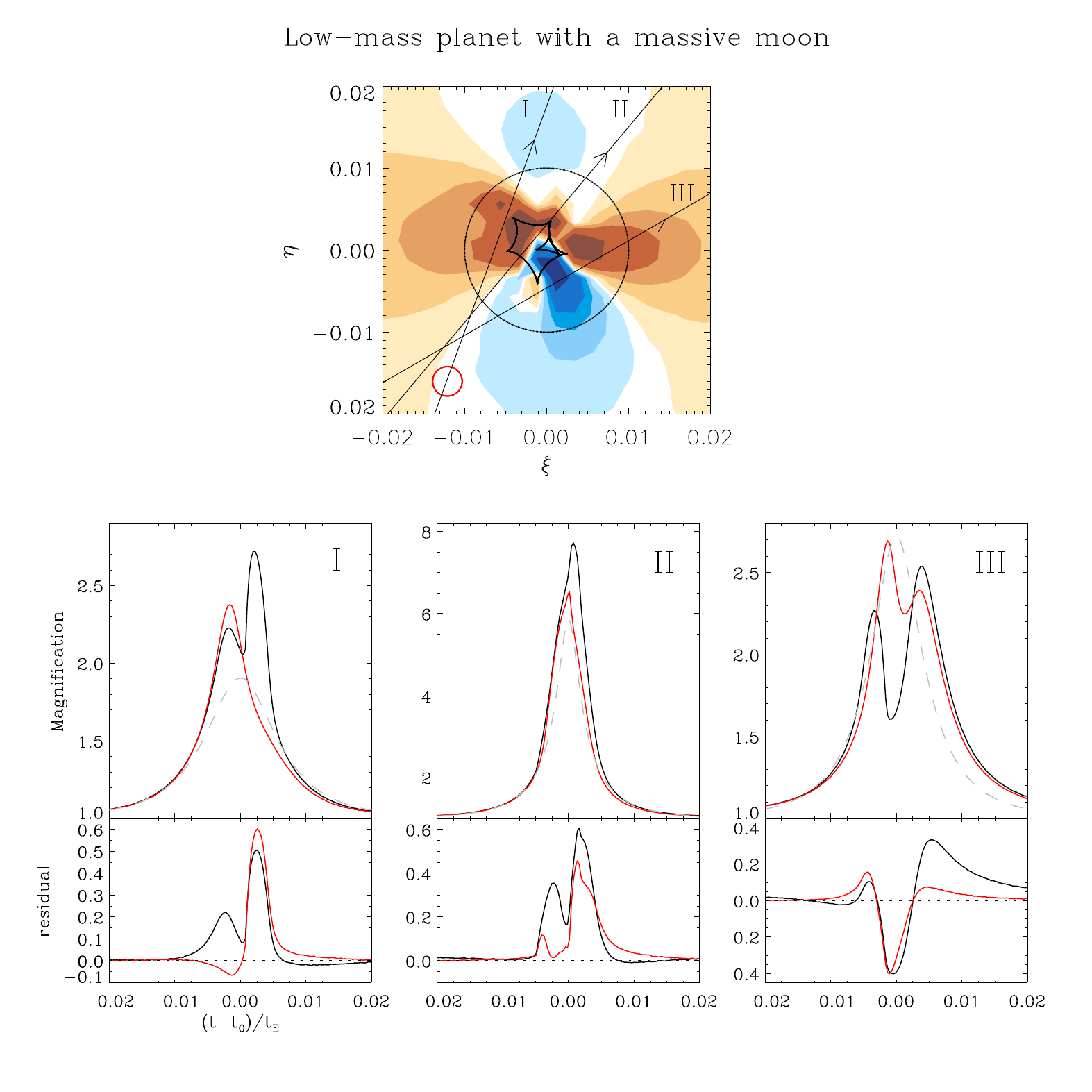}
\caption{\label{fig:six}
Upper panel : Fractional deviation map between a wide separation planetary lens system with a low-mass planet and a massive moon and a single lens system of the planet itself.
The masses of the planet and moon are $10 M_\oplus$ and $M_\oplus$, which correspond to $q_{\rm p} = 1.0\times10^{-4}$ and  $q_{\rm m} = 1.0\times10^{-5}$, respectively, and other lensing parameters are $\spp = 3.0$, $\sm =1.2$, and $\phi = 60^{\circ}$.
The source radius normalized by the assumed typical Einstein radius is $\rho_{\star} = 0.0018$.
Lower panel : Light curves and residuals for the source trajectories presented in the map.
The colored light curves are the same as expressed in Figure 2.
}
\end{figure}

\begin{deluxetable}{ccccccccc}
\tablecaption{Photometric accuracy for the lunar deviation detection.\label{tbl-one}}
\tablewidth{0pt}
\tablehead{
Source Brightness ($I$) && \multicolumn{2}{c}{Photon acquisition rate} && \multicolumn{2}{c}{ Photometric accuracy at $I$ = 18} \\
(mag) && \multicolumn{2}{c}{(photons/s)}  && \multicolumn{2}{c}{($\%$)}\\
\cline{3-9}\\
&& $\delta = 5\%$ & $ \delta = 10\% $ &&  $\delta = 5\% $ & $ \delta = 10\%$ 
}
\startdata
19.0  && 84 & 32  && 0.7 & 1.2 \\
18.5  && 45 & 18  && 1.0 & 1.6 \\
18.0  && 26 & 10  && 1.3 & 2.3
\enddata
\tablecomments{Photon acquisition rate is the photon number per second for an $I = 20.0$ star that is required to detect the lunar deviations of $\delta = 5\%$ and $ \delta = 10\% $ in triple lensing events of each source star. Photometric accuracy is the accuracy at $I = 18.0$ corresponding to each photon acquisition rate. We assume that the radii of the source stars are all $1.0\ R_\odot$.
}
\end{deluxetable}

\acknowledgments
We would like to thank an anonymous referee for making helpful comments and suggestions.
This work was supported by the KASI (Korea Astronomy and Space Science Institute) grant 2016-1-832-01.

\end{document}